\documentclass[amsmath,amssymb,showpacs,10pt]{revtex4}
\usepackage{graphicx}
\usepackage{psfig} 
\newcommand{\s}{\ensuremath{\psi(t,r)}}
\newcommand{\n}{\ensuremath{\nu(t,r)}}
\newcommand{\T}{\ensuremath{\theta}}

\newcommand{\pt}{\ensuremath{p_{\theta}}}
\newcommand{\pr}{\ensuremath{p_r}}

\newcommand{\e}{equation$\;$} 
\newcommand{\M}{\ensuremath{{\cal M}}}

\newcommand{\prz}{\ensuremath{p_{r_0}}}

\newcommand{\ptz}{\ensuremath{p_{\theta_0}}}

\newcommand{\X}{\ensuremath{{\cal X}}}

\newcommand{\ti}{\mbox{$t_{i}$}}

\def\be{\begin{equation}}
\def\eq{\end{equation}}

\begin{document}
\preprint{}
\title{Gravitational collapse from smooth initial data with vanishing radial pressure}
\author{Ashutosh Mahajan}
\email{ashutosh@tifr.res.in}
\author{Rituparno Goswami}
\email{goswami@tifr.res.in}
\author{Pankaj S. Joshi}
\email{psj@tifr.res.in}
\affiliation{Tata Institute for Fundamental Research,\\ Mumbai,
India}

\begin{abstract}
We study here the spherical gravitational collapse assuming
initial data to be necessarily smooth, as motivated by
the requirements based on physical reasonableness. A tangential pressure model 
is constructed and analyzed in order to understand the 
final fate of collapse explicitly in terms of the density and 
pressure parameters at the initial epoch from which the collapse
develops. It is seen that both black holes
and naked singularities are produced as collapse end states even 
when the initial data is smooth. We show that the outcome is
decided entirely in terms of the initial data, as given by 
density, pressure and velocity profiles at the initial epoch, 
from which the collapse evolves. 
\end{abstract} 
\maketitle

\section{Introduction}

An important question in black hole physics is whether it 
would be possible to avoid the naked singularities forming as 
gravitational collapse end states by imposing various possible 
physically reasonable conditions on the collapsing configurations. 
When a sufficiently massive star exhausts its nuclear fuel, it must 
collapse endlessly. From such a perspective gravitational 
collapse has been studied within the framework of general 
relativity extensively for various forms of matter such as dust,
radiation collapse, perfect fluids, massless scalar fields
(see e.g. 
\cite{rev}
and references therein), to find 
that both black holes (BH) and naked singularities 
(NS) result as collapse end states.

Such collapse studies do assume physically reasonable
conditions such as an energy condition, the regularity of initial
data, and such others. It is however, not ruled out that further
conditions, mainly motivated by physical considerations, might
possibly help ruling out naked singularities, and if this turns
out to be the case, this is certainly worth exploring in view of the 
basic importance of this question in the theory and applications of 
black hole physics.

Our purpose here is to examine if both BH/NS phases of 
collapse 
end states would still occur, when in order to make the model  
physically more relevant, we consider only smooth initial profiles
and also include non-zero pressures for the collapsing cloud. 
That is, the initial density and the radial and tangential 
pressures are assumed to be smooth, 
$C^\infty$ functions of the radial co-ordinate $r$. While the Einstein
equations by themselves do not impose such conditions on the 
collapse (where
the metric and other functions are basically required to be $C^2$
differentiable only), smoothness of initial data is frequently 
considered to be desirable, especially in view of the numerical evolutions 
of collapse models related to scalar field collapse
\cite{scalar}. Also, when the 
density and pressures are smooth functions,
there will be, for example, no `cusps' present initially at the center of
the collapsing cloud. Again, while we do not know 
whether the central cores of high density collapsing clouds admit such 
features or not, it may be a good assumption to make on physical 
grounds, and then to examine the consequences in terms of the 
collapse end states.

Our second purpose here is to examine explicitly, how the
collapse outcomes are affected when a non-vanishing pressure is 
taken into account. For the case of vanishing pressures, the outcome of 
dust collapse has been known with or without assumptions 
such as smoothness of initial density distributions
\cite{dust}. 
Here the
initial values of density profile and the velocities of the 
collapsing shells fully determine the collapse end state in terms
of either a black hole or a naked singularity
(see e.g. \cite{dim}).
The situation
is much less explored when pressures are present. 
We examine below a collapse model with a non-vanishing but smooth 
tangential pressure present. It is then seen explicitly
that in fact the smoothness of the initial profiles
cannot restore the Cosmic Censorship Conjecture (CCC) which
forbids occurrence of naked singularities in collapse, and that 
there exist 
non-zero measure of initial data sets which lead to  
outcomes in terms of either BH or NS. The interesting feature about
the model we construct in Section III is that despite the presence of 
pressure, the 
final state is determined entirely in terms of the density
and pressure gradients specified at the initial epoch 
itself, as in the case of dust collapse.

The outline of the paper is as follows.
In Section II we discuss the collapse equations and the regularity 
conditions. In Section III, a tangential pressure model is constructed
and it is 
demonstrated how the given sets of 
initial value parameters such as the initial density and tangential 
pressure values,
decide the collapse final state. The behavior of the tangential 
pressure near singularity is discussed in Section IV, and the 
nature of singularity is examined in Section V. 
Some conclusions are outlined in Section VI.

\section{Einstein Equations, Regularity and Energy conditions}

We use the polar coordinates $(t,r,\theta,\phi)$ 
to write the spherically symmetric metric as,
\begin{equation}
ds^2= -e^{2\n}dt^2 + e^{2\s}dr^2 + R^2(t,r)d\Omega^2
\label{eq:metric}
\end{equation}
where $d\Omega^2$ is the line element on a two-sphere. Also we 
take the energy-momentum tensor to be diagonal for 
the collapsing {\it Type I} matter field 
(that is, the frame is a comoving coordinate 
system) which is given by,
\begin{equation}
T^t_t=-\rho;\; T^r_r=p_r;\; T^\T_\T=T^\phi_\phi=p_\T
\label{eq:setensor}
\end{equation}
This is a fairly general class of matter fields, which includes
various known physical forms of matter
\cite{he}.
The quantities $\rho$, $p_r$ and $p_\T$ are the density, radial 
pressure and the tangential pressure respectively. We take the matter 
field to satisfy {\it weak energy condition}, that is, the energy 
density as measured by any local observer be non-negative, and so 
for any timelike vector $V^i$ we have,
\begin{equation}
T_{ik}V^iV^k\ge0
\end{equation}
This amounts to,
\begin{equation}
\rho\ge0;\; \rho+p_r\ge0;\; \rho+p_\T\ge0
\end{equation}
The dynamical evolution of 
the system is determined by the Einstein equations and for the 
metric (\ref{eq:metric}) these are given as
\be
\rho = \frac{F'}{R^2R'}, \; \; \; \;
  p_{r}=-\frac{\dot{F}}{R^2 \dot{R}}
\label{eq:t6}
\eq
\be
\nu'(\rho+ p_{r})=2(p_{\theta}-p_{r})\frac{R'}{R}-p_{r}'
\label{eq:t7}
\eq
\be
-2 \dot{R'}+R'\frac{\dot{G}}{G}+\dot{R}\frac{H'}{H}=0
\label{eq:t8}
\eq
\be
G-H=1 - \frac{F}{R}
\label{eq:t9}
\eq
\\
where $(\,\dot{}\,)$ and $(')$ represent partial derivative with respect 
to $t$ and $r$ respectively and,
\be
G(r,t)=e^{-2\psi}(R')^2, \; \; H(r,t)=e^{-2\nu}\dot{R}^2
\eq       
Here $F(r,t)$ is an arbitrary function, and in spherically
symmetric spacetimes, it is called the mass function of the collapsing 
cloud, which may be interpreted to give the total mass within the shell of 
comoving radius $r$.
In order to preserve the regularity at the initial epoch, 
$F(t_i,0)=0$, that is, the mass function should vanish at 
the center of the cloud.
It can be seen from the equation ($\ref{eq:t6}$) that the density 
of the matter blows 
up when $R=0$ or $R'=0$. Here the case $R'=0$ corresponds to the {\it 
shell-crossing} singularities. However, it is widely believed (see e.g.
\cite{shell})
that these 
singularities can be possibly removed from the spacetime
as they are typically gravitationally weak. 
Hence, we shall consider here only the {shell-focusing} 
singularities which occur at $R=0$, where the physical
radius of all the matter shells goes to a zero value. 
Now let us use the scaling independence of the coordinate $r$ to write,
\begin{equation}
R(t,r)=rv(t,r)
\label{eq:R}
\end{equation}
and we have,
\begin{eqnarray}
v(t_i,r)=1; & v(t_s(r),r)=0; & \dot{v}<0
\label{eq:v}
\end{eqnarray}
where $t_i$ and $t_s$ stand for the initial and the singular   
epochs respectively.  The condition $\dot{v}<0$ signifies
that we are dealing with collapse situations only.
This means we scale the radial coordinate $r$ in such a way that at 
the initial epoch $R=r$, and at the singularity, $R=0$. 
The advantage that the introduction of this new variable
$v$ offers is that the regular center at $r=0$ (where we also
have $R=0$) is now distinguished
from the genuine singularity at $R=0$ in that, we now have
$v=1$ at the initial epoch, and $v=0$ at the singular
epoch $R=0$, but at all other epochs in-between $v$ has a 
non-zero finite value for all values of $r$.

From the point of view of dynamic evolution of initial data, at 
the initial epoch $t=t_i$, we now have five arbitrary functions of 
coordinate $r$ as given by,
\begin{eqnarray}
\nu(t_i,r)=\nu_0(r); & \psi(t_i,r)=\psi_0(r); &\rho(t_i,r)=\rho_0(r)\nonumber\\
p_r(t_i,r)=p_{r_0}(r); &\pt(t_i,r)=p_{\T_0}(r)
\label{eq:init}
\end{eqnarray}
We note that all the initial data represented by the equation 
(\ref{eq:init}) are not mutually independent, as from 
equation (\ref{eq:t7}) we get,
\begin{equation}
\nu_0(r)=\int_0^r\left(\frac{2(p_{\T_0}-p_{r_0})}{r(\rho_0+p_{r_0})}
-\frac{p_{r_0}'}{\rho_0+p_{r_0}}\right)dr
\label{eq:nu0}
\end{equation}
Now, to preserve the regularity and smoothness of the initial 
data let us make some assumptions about the initial pressures 
at the regular center $r=0$.
Let the gradients of pressures vanish at the center, that is, 
$\prz'(0)=\ptz'(0)=0$.
The difference between radial and tangential pressures at the center
should also vanish, i.e. $\prz(0)-\ptz(0)=0$.
With these physical assumptions, and from equation (\ref{eq:nu0}), it is 
evident that 
$\nu_0(r)$ has the form,
\begin{equation}
\nu_0(r)=r^2g(r)
\label{eq:nu0form}
\end{equation} 
where $g(r)$ is at least a $C^1$ function of $r$ at $r=0$, 
and at least a $C^2$ function for $r>0$.
Hence we see that we have a total of five field equations with 
seven unknowns, $\rho$, $p_r$, $\pt$, $\psi$, $\nu$, $R$, and $F$, 
giving us the freedom of choice of two free functions. Selection 
of these functions, subject to the given initial data and weak 
energy condition, determines the matter distribution and metric 
of the space-time and thus leads to a particular dynamical
collapse evolution of the initial data. For collapsing 
models we have $\dot R<0$.

\section{A Collapse model with vanishing radial pressure and non-vanishing
tangential pressure}

The spherically symmetric collapse models, where the radial 
pressure is taken to be vanishing, but the tangential pressure could
be non-zero have been studied in some detail over past years 
\cite{tan}.
The
main motivation here was, if pressures are introduced within a collapsing
cloud, one would like to understand how the situation differed from the
dust collapse, especially in terms of formation of black holes and
naked singularities as collapse end states. The Einstein cluster
\cite{ein}
is an example of such a cloud where tangential stresses are present.
This is a spherically symmetric cluster of rotating particles where
the motion of the particles is sustained by an angular momentum which
has an average effect of creating a non-zero tangential stress 
within the cloud.

What is clear now as we study the models with non-vanishing 
tangential pressures is that both BH/NS phases do develop as collapse
end states, that is, the naked singularities of dust collapse 
do not go away with the introduction of tangential pressure. In that 
sense, these are stable to introduction of pressure. While these are existence results, we would like to understand how the nature of the final state will depend on the nature of the initial data given (see also \cite{einclu} for the case of Einstein cluster).

In the following, we construct an explicit example of a collapse
model with a non-vanishing tangential pressure, and it is shown
that the nature of the singularity in this case can be determined
explicitly in terms of the initial values of the density and
pressure parameters, as given at the initial surface from which
the collapse develops. This is then parallel to the dust
case where such results are available.

As pointed out in the previous section, we choose the 
two allowed free functions, $F(t,r)$ and $\n$ 
in the following way,
\begin{equation}
F(t,r)=r^3\M (r)
\label{eq:mass}
\end{equation}
that is, we choose the class of mass functions where $\M$
is a function of the radial coordinate $r$ only. Also we choose,
\begin{equation}
\n=c(t)+\nu_0(R)
\label{eq:nu}
\end{equation}

Our purpose here is to examine 
to what extent the initial data determine collapse final fate, 
and hence we desire to construct here explicitly a specific model of 
collapse with 
non-vanishing tangential pressure in which case we will show that both BH 
and     
NS are possible as end states from smooth initial data, and that 
it is the initial data set which determines the final outcome.
This is definitely a generalization over the dust models where initial    
profiles fully determine the collapse end state. Eq. (16), in
spite of being a strong assumption, enables us in doing this.

Now putting equation (\ref{eq:mass}) in equation (\ref{eq:t6}), we get,
\begin{eqnarray}
\rho=\frac{3\M+r\left[\M_{,r}\right]}{v^2(v+rv')};&&\pr=0
\label{eq:rho} 
\end{eqnarray}
Thus we see that the given choice of the mass function ensures
the radial pressure to vanish identically throughout the collapse.
Also, it is evident that the density at the initial epoch is 
given by,
\begin{equation}
\rho_0(r)=3\M(r)+r\M(r)_{,r}
\label{eq:rho0}
\end{equation}
It is clear that, in general, as $v\rightarrow 0$, 
$\rho\rightarrow\infty$.
Thus the density blows up at the singularity $R=0$
which will be a curvature singularity as expected.
Also using \e (\ref{eq:nu}) in \e (\ref{eq:t8}), we have,
\begin{equation}
G(t,r)=b(r)e^{2\nu_0(R)}
\label{eq:G}
\end{equation}
Here $b(r)$ is another arbitrary function of $r$. 
In correspondence with the dust models, we can write,
\begin{equation}
b(r)=1+r^2b_0(r)
\label{eq:veldist}
\end{equation}
where $b_0(r)$ is the energy distribution function for the 
collapsing shells.

Now let us consider a smooth initial data, {\it i.e.} the initial 
density, pressure, and energy distributions are expressed as  
only even powers of $r$. 
\be
\rho(\ti,r)=\rho_{00}+\rho_{2}r^2 +\rho_{4}r^4+\cdots
\eq
\be
p_{\theta}(\ti,r)=p_{\theta_2}r^2 +p_{\theta_4}r^4+\cdots 
\eq
\be
b_0(r)=b_{00}+b_{02}r^2+\cdots
\eq 
With the above form of smooth initial data to evolve the collapse,
we can integrate the equation (\ref{eq:nu0})
and get,
\be
\nu_{0}(R)=p_{\theta_2} R^2 +\frac{(p_{\theta_4}-\rho_{2}p_{\theta_2})}
{2} R^4+\cdots
\label{eq:nu1}
\eq

Using now \e(\ref{eq:nu}) in \e(\ref{eq:t7}), we get
\begin{equation}
2\pt=R\nu_{,R}\rho
\label{eq:ptheta}
\end{equation}

Finally, using equations 
(\ref{eq:mass}),(\ref{eq:nu}) and (\ref{eq:G}) in \e(\ref{eq:t9}), we have,
\begin{equation}
\sqrt{R}\dot{R}=-a(t)e^{\nu_0(R)}\sqrt{(1+r^2b_0)Re^{2\nu_0}-R+r^3\M}
\label{eq:collapse}
\end{equation}
Here $a(t)$ is a function of time. By a suitable scaling of the time 
coordinate, 
we can always make $a(t)=1$. The negative sign is due to the fact that 
$\dot{R}<0$, which is the collapsing cloud condition.\\

Let us define a function $h(R)$ as,
\begin{equation}
h(R)=\frac{e^{2\nu_0(R)}-1}{R^2}=2g(R)+{\cal O}(R^2)
\label{eq:h}
\end{equation}
Using \e(\ref{eq:h}) in \e(\ref{eq:collapse}), we have after simplification,
\begin{equation}
\sqrt{v}\dot{v}=-\sqrt{e^{4\nu_0}vb_0+e^{2\nu_0}\left(v^3h(rv)+\M\right)}
\label{eq:collapse1}
\end{equation}
Integrating the above equation, we get,
\begin{equation}
t(v,r)=\int_v^1\frac{\sqrt{v}dv}{\sqrt{e^{4\nu_0}vb_0+e^{2\nu_0}
\left(v^3h(rv)+\M\right)}}
\label{eq:scurve1}
\end{equation}
The time of singularity for a shell at a comoving coordinate 
radius $r$ is the time when the 
physical radius $R(r,t)$ becomes zero. 
The shells collapse consecutively, that is one after the other to the 
center as there are no 
shell-crossings.
Taylor expanding the above function around $r=0$, we get, 
\be 
t(v,r)=t(v,0)\;+\left.r\;\frac{d t(v,r)}{dr}\right|_{r=0}  
+\left.\frac{r^2}{2!}\;\frac{d^{2}t(v,r)}{d^2{r}^2}\right|_{r=0}
\label{eq:scurve2}
\eq
Let us denote, 
\be
\X_{n}(v)=\left.\frac{{d} ^{n} t(v,r)}{{d} r^{n}}\right|_{r=0}  
\eq
As we have taken the initial data with only even powers of $r$, 
the first derivatives
of the functions appearing in above equations vanish at $r=0$, hence
we have,
\be
\X_{1}(v)=0
\eq
Now we can express the next coefficient $\X_{2}$ as,
\be
\X_{2}(v)=-\int_{v}^{1}\frac{\sqrt{v}dv[\Delta v^5 + \frac{2}{3}p_{\theta_{2}}
v^2+2b_{02}v+\frac{\rho_{2}}{5}]}{(hv^3+b_{00}v+ M)^{\frac{3}{2}}}
\label{eq:scurve5}
\eq
where,
\be
\Delta=6p_{\theta_{2}}^2+8b_{02}p_{\theta_{2}}-\rho_{2}p_{\theta_{2}}
\eq

When $\nu=0$, it can be seen from equation (\ref{eq:ptheta}) that 
the tangential pressure vanishes and the model becomes 
like dust. The time
$t(v,r)$ and coefficient $\X_{2}$ for the dust case are given as

\begin{equation}
t^{d}(v,r)=\int_v^1\frac{\sqrt{v}dv}{\sqrt{vb_0+\left(v^3h(rv)+\M\right)}}
\label{eq:scurved1}
\end{equation}

\be
\X_{2}^{d}(v)=-\int_{v}^{1}\frac{\sqrt{v}dv[2b_{02}v+\frac{\rho_{2}}{5}]}
{(b_{00}v+\M)^{\frac{3}{2}}}
\label{eq:scurved5}
\eq
The important point to note here is that any given initial 
profile given in terms of the density and pressure
values at the initial epoch completely determines the 
function $\X_{2}(v)$.

In general, for any constant $v$ surface we have,
\be
\sqrt{v}v'=-\sqrt{v}\dot{v}\frac{dt}{dr}
\eq
We see that the time taken for the central shell to reach the singularity 
is given as
\begin{equation}
t_{s_0}=\int_0^1\frac{\sqrt{v}dv}{\sqrt{vb_0+v^3h(0)+\M_0}}
\label{eq:scurve3}
\end{equation}

From the above equation it is clear that for $t_{s_0}$ to be defined,
\begin{equation}
b_{00}v+h(0)v^3+\M_0>0
\label{eq:constraint}
\end{equation}
Hence the time taken for other shells to reach the singularity 
can be given by the expression,
\begin{equation}
t_s(r)=t_{s_0}+\frac{1}{2}r^2\X_2(0)+{\cal O}(r^3)
\label{eq:scurve4}
\end{equation}

\section{Behavior of tangential pressure near the singularity}

From the equations (\ref{eq:nu1}) and  (\ref{eq:ptheta}), it is 
clear that near the center the tangential pressure behaves as,
\begin{equation}
\pt\sim\frac{r^2}{R'}
\label{ptheta2}
\end{equation}
If we wish to calculate the limit of $\pt$ along any curve approaching
the central singularity then in $(v,r)$ plane along all these curve
$v\rightarrow0$ as $r\rightarrow0$. Then in $(v,r)$ plane the equation
of all these radial curves near the center should have the form,
\begin{equation}
v=k_1r^\alpha\;\;\;\;(\alpha>0)
\label{curve1}
\end{equation}
Rewriting the equation of these curves in $(t,r)$ plane we have,
\begin{equation}
\frac{dt}{dr}=\frac{\partial t}{\partial r}+\frac{\partial t}{\partial v}
\left(\frac{dt}{dr}\right)_{\rm along\;the\;curve}
\label{curve2}
\end{equation}
Now using (\ref{eq:scurve2}) and calculating the limits on the given curve
we finally get in $(t,r)$ plane that the above {\it ingoing} 
curve can be written as,
\begin{equation}
t_{s_0}-t=r^2\X_2(0)+k_2r^{(3\alpha/2)}
\label{curve3}
\end{equation}
Now if $\X_2(0)\neq0$ it can be easily shown that the limit of $\pt$
in (\ref{ptheta2}) does not blow up, and the system behaves like
dust essentially near the central singularity. In case of $\X_2(0)=0$ 
it can be shown that for the values of $\alpha>2$ the pressure blows up.
However from the metric it can be calculated that if such curves are
to be timelike then strictly $\alpha<2$. Hence we see that there exist
no timelike radial curve along which the tangential pressure blows up
at the central singularity. However there are a large class of spacelike
radial directions, described by $\X_2(0)=0$ and $\alpha>2$, 
along which it blows up at the central singularity.\\

\par
Alternatively, we can also look at the solution close to the center. 
In the limit $R\to 0$ (for the zero initial velocity profiles),
the equation  \e(\ref{eq:collapse}) takes the form,

\begin{equation}
R\dot{R}^{2}= 2\,p_{\theta_2}r^3\M(r)\, R^{2}+r^3\M(r)
\label{eq:solncenter}
\end{equation}

which can be integrated as

\be
\frac{1}{\sqrt{c_2}}R^{\frac{3}{2}}
F[\frac{1}{2},\frac{3}{4},\frac{7}{4},-\frac{c_1}{c_2}R^{2}]= t-t_{s}(r)
\eq
 
where $F$ is hypergeometric function,
$c_{1}=\frac{9}{2}\,p_{\theta_2}\,r^{3}M(r)$ and 
$c_{2}= \frac{9}{4}r^{3}\,M(r)$. \\

Expanding the hypergeometric function $F$ and keeping only the 
leading order term, we get the solution for $R$ as

\be
R\,=\, \left(\frac{3}{2}\right)^{\frac{2}{3}}\,r\,M(r)^{\frac{1}{3}}\,(t-t_{s}(r))
^{\frac{2}{3}}
 \eq  
\par
In the equation (\ref{eq:ptheta}), if we expand $\nu$ and $\rho$ we see that
near the central singularity in the limit $R\to 0$ the 
pressure $p_{\theta}$ goes as $(p_{\theta2}\rho_{00}) (r^2/R')$.
Now, in $(R,r)$ plane, along a curve $R=kr^b$ the quantity $R'$ can 
be calculated 
from the above equation, and when we approach the central singularity 
we see that the tangential pressure blows up for $b>3$.\\

We see that close to the center, this curve can be timelike 
only if $b<3$. Thus 
for the curve $R=ar^b$, $p_{\theta}$ blows up only along spacelike directions.
A similar analysis was carried out for the curves like  $t-t_{so}=\alpha \, 
r^{\beta}$ and it was found that $p_{\theta}$ blows up for the 
value of $\beta$ for which the curve is spacelike. \\

\section{Nature of the singularity}

We need to determine now when there will be
families of future directed outgoing null geodesics coming out
of the singularity and when there will be none. In the case when
such families do exist which terminate in the past at the singularity,
and which could reach outside observers, then the singularity will
be visible. In the case otherwise it is hidden within the
black hole. Another way to look at this is through the apparent horizon
and formation of trapped surfaces in the spacetime. As the
collapse evolves, if the trapped surfaces form well in advance to 
the formation of the singularity, then the same will be covered.
On the other hand, if the trapped surface formation is sufficiently 
delayed during the collapse then the singularity may be naked.
The apparent horizon within the collapsing cloud is given by the \e,
$R/F=1$, which gives the boundary of the trapped surface region 
of the space-time. If the neighborhood of the center gets 
trapped earlier than the singularity, then it is covered, 
otherwise it is naked with non-spacelike future directed 
trajectories escaping from it.

In order to consider the possibility of existence 
of such families, and to examine the nature of the singularity 
occurring at $R=0, r=0$ in this model, let us consider the 
outgoing null geodesic equation which is given by,
\begin{equation}
\frac{dt}{dr}=e^{\psi-\nu}
\label{eq:null1}
\end{equation}
We now use here a method 
which is 
similar to that given in
\cite{dim}.
The singularity curve is given by $v(t_s(r),r)=0$, which corresponds 
to $R(t_s(r),r)=0$. Therefore, if we have any future directed 
outgoing null 
geodesics terminating in the past at the singularity, we must have 
$R\rightarrow 0$ as $t\rightarrow t_s$ along the same. 
Now writing \e 
(\ref{eq:null1}) 
in terms of variables $(u=r^\alpha,R)$, we have,
\begin{equation}
\frac{dR}{du}=\frac{1}{\alpha}r^{-(\alpha-1)}R'\left[1+\frac{\dot{R}}{R'}
e^{\psi-\nu}\right]
\label{eq:null2}
\end{equation}

Now in order to get tangent to the null geodesic in the $(R,u)$ plane,
we choose a particular value of $\alpha$ such that the geodesic equation 
is expressed only in terms of $\left(\frac{R}{u}\right)$. 
A specific value of alpha is to be chosen which enables us to calculate
the proper limits at the central singularity.
For example, 
for $\X_1(0)\ne0$ case,
we can choose $\alpha=\frac{5}{3}$ and using \e (\ref{eq:t9}), 
(and considering that $\dot{R}<0$), we get,
\begin{equation}
\frac{dR}{du}=\frac{3}{5}\left(\frac{R}{u}+\frac{\sqrt{\M_0}\X_1(0)}
{\sqrt{\frac{R}{u}}}\right)\left(\frac{1-\frac{F}{R}}{\sqrt{G}[\sqrt{G}
+\sqrt{H}]}\right)
\label{eq:null3}
\end{equation}
In the tangential pressure collapse model discussed in the 
previous section we have $\X_1(0)=0$, and hence
we choose $\alpha=\frac{7}{3}$ so that when in limit $r\rightarrow 0, 
t\rightarrow t_{s}$ we get the value of tangent to null geodesic in 
the $(R,u)$ plane,
\be
\frac{dR}{du}=\frac{3}{7}\left(\frac{R}{u}+\frac{\sqrt{M_{0}}\X_{2}(0)}
{\sqrt{\frac{R}{u}}}\right)\frac{(1-\frac{F}{R})}{\sqrt{G}(\sqrt{G} 
+\sqrt{H})}
\label{eq:null4}
\eq

Now note that for any point with $r>0$ on the singularity curve 
$t_s(r)$, we have $R\to 0$ whereas $F$ (interpreted as mass of the
object within the comoving radius $r$) tends to a finite positive
value once the energy conditions are satisfied. Under the situation,
the term $F/R$ diverges in the above equation, and all such points 
on the singularity curve will be covered as there will be no outgoing
null geodesics from such points.

Hence we need to examine the central singularity at $r=0,R=0$
to determine if it is visible or not. That is, we need to determine
if there are any solutions existing to the outgoing null geodesics 
equation, which terminate in the past at the singularity and in future
go to a faraway observer, and 
if so under what conditions these exist. Note 
that if any outgoing 
null geodesics terminate at the singularity in the past then along the 
same, in the limit as $r\rightarrow 0, t\rightarrow t_{s}$ 
we then have from equation (26) $\dot{R} =0$, therefore $H=0$ and  $G=1$ 
in this limit as $F/R$ vanishes.
Let now $x_{0}$ be the tangent to the null geodesics in $(R,u)$ plane, 
at the central singularity, then it is given by,
\begin{equation}
x_0=\lim_{t\rightarrow t_s}\lim_{r\rightarrow 0} 
\frac{R}{u}=\left.\frac{dR}{du}\right|_{t\rightarrow t_s;r\rightarrow 0}
\end{equation}
Using \e (\ref{eq:null4}), we get,
\begin{equation}
x_0^{\frac{3}{2}}=\frac{7}{4}\sqrt{\M_0}\X_2(0)
\end{equation}
In the $(R,u)$ plane, the null geodesic equation will be,
\be
R=x_0u
\eq
while in the $(t,r)$ plane, the null geodesic equation near 
the singularity will be,
\begin{equation}
t-t_s(0)=x_0r^{\frac{7}{3}}
\end{equation}
It follows that if $\X_2(0)>0$, then that implies that $x_0>0$, 
and we then have radially 
outgoing null geodesics coming out from the singularity, making the 
central singularity to be a visible one. On the other hand, if 
$\X_2(0)\le0$, we will have a 
black hole solution.

We have, however, already seen in equation 
(\ref{eq:scurve5}), 
that the value of $\X_2(0)$ entirely depends upon the initial 
density and velocity profiles. 
Therefore, given any density and tangential pressure profiles 
of the collapsing matter, we can always choose an energy profile 
so that the end state of 
the collapse would be either a naked singularity or a black hole 
and vice-versa.

These features come out clearly in the Figures 1-3,
where it is examined how $\X_2$ depends upon the various 
initial
data values chosen. Here the function
$\X_2(0)$ is plotted with different values of initial density, tangential
pressure and energy profiles. It is seen that there exist distinct regions
in the initial data space where a certain region evolves to a naked 
singularity outcome, whereas the other part of the region evolves to a 
black hole.
In the figures the red colored surface shows positive values of $\chi_{2}$.
Whenever the initial data values lie in this region, null geodesics from 
the center come out of the central singularity, as a result
creating the presence of a naked singularity as the collapse outcome.
The green colored region represents black hole. The phase separation 
is clearly seen.
It is to be noted that in case of a naked singularity, the 
singularity 
curve at the center, as given by equation (\ref{eq:scurve4}), is an 
increasing function of $r$ as in that case $X_2(0)>0$, whereas a 
black hole solution gives a decreasing or constant curve for 
shells as the coordinate $r$ increases.

These plots also bring out another important feature arising
in collapse, which is that the initial data sets which give rise to
either a naked singularity or a black hole as final end state are
open in the space of initial data from which the collapse evolves.

We note that even though the model considered here has non-zero
pressure, this is subject to a strong assumption given by Eq. (16).
This is equivalent to the ansatz $\nu(t,r) = \nu_0(R)$. A feature that
the plots given here bring out is, in the space of initial data that
is defined under these restrictions the BH/NS outcomes are open sets.
Though restricted as above, this scenario is more general as compared
to dust collapse, in that pressure is included. 

This of course does not mean that, in general, naked singularities
would generically develop in any given collapse scenario.
However, it is to be noted that the issues of genericity and stability
are rather involved in Einstein's gravity, and are not so well-defined.
Hence these must always be studied within the context of the limited
models under consideration. It is to be hoped that such considerations
will possibly throw some light on the nature of cosmic censorship statement
which one may evolve eventually.

\begin{figure}[ht!!]
\centering
\includegraphics[width=6.5cm,angle=-90]{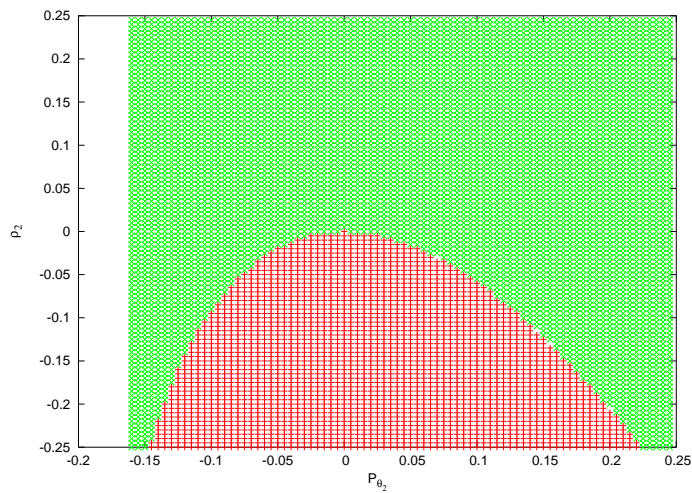}
\caption{Illustration of the dependence of end state of the collapse on 
initial density $\rho_{0}$ and tangential pressure $p_{\theta_{0}}$ for $b_{02}=0$. The red coloured region shows the
initial data space which goes to NS, while the green region evolves
to BH.} 
\end{figure}
\begin{figure}[ht!!]
\centering
\includegraphics[width = 6.5cm,angle=-90]{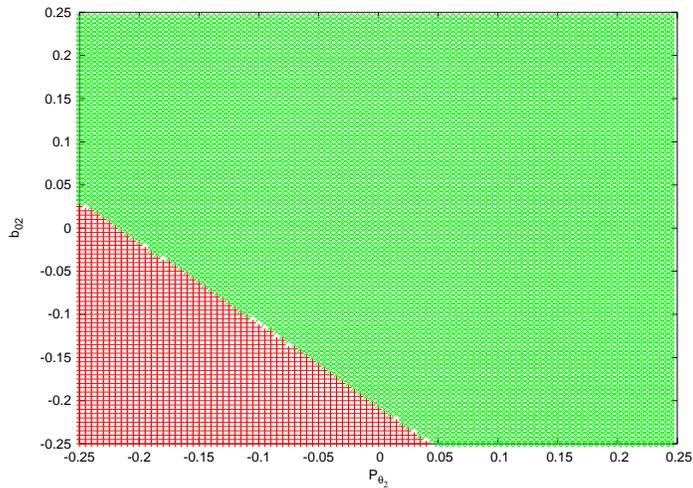}
\caption{Illustration of the dependence of end state of the collapse on
initial tangential pressure $p_{\theta_{0}}$ and 
energy profile $b_{02}$ for
$\rho_{2}=1$. The red coloured region shows the
initial data space which goes to NS, while the green region evolves
to BH.} 
\end{figure}
\begin{figure}[htb!!]
\centering
\includegraphics[width = 6.5cm,angle=-90]{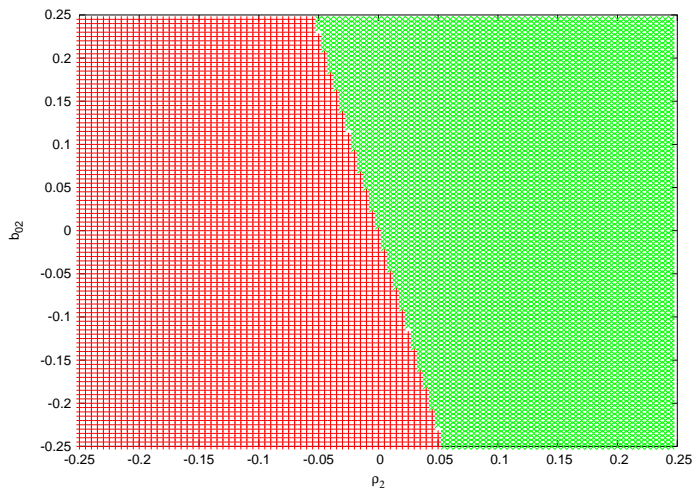}
\caption{Illustration of the dependence of end state of the collapse on
initial density $\rho_{0}$ and energy profile $b_{02}$ for
$p_{\theta_{2}}=0$. The red coloured region shows the
initial data space which goes to NS, while the green region evolves
to BH.}
\end{figure}

\section{Conclusions}

We have considered here gravitational collapse from a smooth
initial data. We have also included the presence of pressure while
developing the collapse models. In this sense, the models here may
be considered closer to being physically realistic, when compared to, for
example, the dust collapse where the pressures are completely
neglected. Some conclusions and remarks are summarized below. \\

1. We see that for the model considered here in Section III, where vanishing
radial pressure is assumed and an additional 
assumption (16) is made, both BH/NS phases develop as collapse end 
states. The important feature that we observe is the outcome is
fully determined in terms of the initial profiles for density,
pressures, and energy functions as given at the initial surface from 
which the collapse
develops. This is completely due to the strong assumption (16).
The point that is made here is, we have a construction or an example 
to show that naked 
singularities can still develop even when densities and 
pressures are restricted 
to be necessarily smooth. Secondly, as seen in the present case, 
the initial data 
alone could determine the final fate even when 
non-zero pressures are present in the collapsing cloud. 
It remains to be examined if this would hold in still wider
classes of models other than those considered here.\\

2. We find in the above case, that there is an open  
initial data space which evolves to a naked singularity, and 
the same is true for a black hole end state as well. 
In this sense both these collapse outcomes are generic,
within the framework of the models considered here.\\

3. The current analysis shows that
only the requirements of smoothness of initial data, and the 
inclusion of pressures, are not sufficient by themselves to rule
out naked singularities. As we see here, smooth
initial data does give rise to NS.\\

4. Through out we have imposed the smoothness condition,
however, it is worth noting that the data is required to be smooth 
only {\it initially}. The subsequent dynamical evolution of these initial
matter profiles is then fully governed by the Einstein equations, and
hence at later epochs during the collapse the density and pressure
profiles may or may not continue to be smooth, and could very well
develop non-analytic behaviour. It may be worth examining, especially
when pressures are included, as to how the smoothness condition
may be affected as the collapse evolves.\\

5. It is important to note
that the analysis presented here only considered locally 
naked singularities.
The null geodesics coming out from the singularity can in principle 
go to infinity, ({\it i.e} the singularity can be globally visible), 
or it could also happen that the trajectories could  again fall back to 
the singularity. 
This in fact depends on the global behavior (i.e. for large values
of $r$) of the functions concerned. In the case when a singularity
is locally naked, one can always choose the global behavior of the 
rest of the dynamical functions in such a way that the singularity becomes 
globally visible.\\

\end{document}